# Spreadsheets Grow Up: Three Spreadsheet Engineering Methodologies for Large Financial Planning Models

Thomas A. Grossman
University of San Francisco, School of Business & Professional Studies,
San Francisco CA  94117-1045
tagrossman@usfca.edu

Özgür Özlük
Decision Sciences Department, College of Business, SFSU
San Francisco, CA 94132
ozgur@sfsu.edu

ABSTRACT

Many large financial planning models are written in a spreadsheet programming language (usually Microsoft Excel) and deployed as a spreadsheet application. Three groups, FAST Alliance, Operis Group, and BPM Analytics (under the name "Spreadsheet Standards Review Board") have independently promulgated standardized processes for efficiently building such models. These spreadsheet engineering methodologies provide detailed guidance on design, construction process, and quality control. We summarize and compare these methodologies.  They share many design practices, and standardized, mechanistic procedures to construct spreadsheets. We learned that a written book or standards document is by itself insufficient to understand a methodology. These methodologies represent a professionalization of spreadsheet programming, and can provide a means to debug a spreadsheet that contains errors.  We find credible the assertion that these spreadsheet engineering methodologies provide enhanced productivity, accuracy and maintainability for large financial planning models.

1. INTRODUCTION

There have been many suggestions in the literature about how to create effective spreadsheets. One challenge is that the universe of spreadsheets is so large that a one-size-fits-all approach will not be useful.  What is needed is a "spreadsheet engineering methodology" (Grossman 2002) that applies to a particular well-defined problem domain, and provides detailed, integrated guidance on design, construction, and documentation of spreadsheet software.

"Spreadsheet engineering" is software engineering applied to spreadsheets. Software engineering methodologies for traditional programming languages are widely researched and used. Universities offer classes and degrees in software engineering; there are many books and theoretical and empirical research articles. The American government even sponsors a highly-respected Software Engineering Institute at Carnegie-Mellon University.

One must recognize that the spreadsheet is a powerful computer programming language. In computer science terms, the spreadsheet is a "fourth-generation language" (FORTRAN and C are third-generation), is an example of a "rapid development language" (which has important benefits and disbenefits), and is an "integrated development environment (that helps programmers create application software by automating or simplifying common tasks); see Grossman, Mehrotra and Özlük (2007) and Grossman (2006). Writing spreadsheet cell formulas is indeed computer programming, and computer programmers need to use software engineering methodologies.

In recent years, three specialist spreadsheet development organizations have each independently developed a spreadsheet engineering methodology. These groups are FAST Alliance, Operis Group, and BPM Analytics (under the name "Spreadsheet Standards Review Board") These methodologies focus on writing spreadsheet application software (i.e., a spreadsheet written by professional programmers for use by others). They contain closely integrated design and process recommendations and a limited domain that set them apart from all previous efforts.

These three spreadsheet engineering methodologies seek consistent high quality and productivity and to develop spreadsheets that are easy to maintain. Each methodology has been used in the organizations' own practice for several years and disseminated in commercial training courses.

In this paper, we summarize and compare three spreadsheet engineering methodologies. We discuss their effectiveness and raise issues for future research.

## 2. OUR APPROACH

We read the published documentation for each methodology. In each case we realized that the documentation was not sufficient to adequately understand the methodology. We contacted each vendor who provided additional information as described below. We then sent each vendor a draft of our summary of their approach and incorporated some of their feedback.

For the FAST methodology, we read the recently published standard document (FAST 2010). In December 2009 we were able to visit the offices of F1F9 in Delhi where we watched an introductory training presentation by staff, observed team spreadsheet development, explored FAST spreadsheets and discussed the philosophy and implementation with staff members. F1F9 also provided a spreadsheet template for our review, and we watched an introductory Webinar on their spreadsheet creation practices.

For the Operis methodology, we read Swan (2008) that describes the approach. Operis provided access to extensive, detailed proprietary training materials. These materials, including a "build" that shows across a series of spreadsheets how a model changes from start to finish, were very helpful and gave us confidence that we understand the central principles and practices of this approach.

For the SSRB approach, we read two standards document (SSRB 2005a and 2005b). We contacted SSRB and they provided a helpful demonstration of their approach, which relies heavily on a powerful spreadsheet add-in. We have a lower level of confidence in our understanding of the SSRB approach.

For each methodology we learned more than we can include in a brief survey paper. We endeavored to present accurately the most important elements of each methodology. We emphasize that any errors of omission or commission are the sole responsibility of the authors.

## 3. LITERATURE SURVEY

There have been a few sets of detailed spreadsheet engineering recommendations for creating general spreadsheet models. The most complete are Read and Batson (1999) and Tennent and

Friend (2001). Powell and Baker (2009, chapter 5) and Raffensperger (2001) made widely-read contributions. These approaches represent a thoughtful effort to address an important problem, and provide useful practices. However, when compared to the spreadsheet engineering methodologies considered in this paper, they are smaller in scope, have more general or poorly-defined domain, lack integration between design decisions and the practices used to construct the design, and/or do not fully address techniques for increasing the speed of development which is critically important to professional financial spreadsheet programmers.

We note that many banks and consulting firms have in-house spreadsheet development courses. For example, one consulting firm, whose name we cannot disclose, shared with us a detailed set of practices that rise to the level of a methodology. However, this methodology is proprietary and cannot be disclosed. It is possible that many such proprietary methodologies exist.

In the next three sections, we summarize the three spreadsheet engineering methodologies for large financial models and highlight what seem to be their most important aspects.

## 4. THE FAST SPREADSHEET ENGINEERING METHODOLOGY

FAST is promulgated by the FAST Modeling Alliance, led by members of F1F9 (India) Pvt Ltd and Financial Mechanics Ltd who also offer training courses in the methodology. A handful of companies are "signatories" to the standard, and there is a provision for providing feedback and suggestions. The FAST standard (FAST 2010) was first published in March 2010.

FAST takes the perspective that the spreadsheet is the *lingua franca* of financial communication and that a financial spreadsheet model is a tool for communication and collaboration as well as calculation. Therefore clarity, readability, and standardization are essential. A common column structure is used on all worksheets, with several narrow columns at the left that are cleverly used to provide section and sub-section headings and aid keyboard navigation. Important design standards include: all inputs to a calculation are placed adjacent to the calculation; inputs reference the original source; calculation modules are kept to minimum size. Important coding standards include: spaces are placed between each element in a formula (as spaces are used between words in a sentence); cell formulas are short; formulas across a row are the same (right to left consistency); and all rows are labeled and have units.

The FAST design relies upon automation in the form of keyboard shortcuts for rapidly constructing and navigating an Excel model. This allows a knowledgeable reader to quickly traverse the spreadsheet to understand the program and verify the formulas, yet quickly and easily return to the calculation being examined. The FAST design leverages the capabilities of the Excel auditing tools to illuminate model structure.

FAST utilizes what we would call "extreme modularity". Multiple worksheets are used for calculations, grouped by function. Any calculation is done in a module of code which FAST calls a "calculation block". With few exceptions, each block contains a single calculation. The top rows of the block contain all the inputs required for the calculation (called "ingredients") and nothing else. In general, all reference cells are echoed to the calculation block so that no distant cells are referenced. Each cell in the input rows points to the original value of that input. This adjacency of labeled inputs reduces the readability benefit of Excel Names and FAST generally avoids Names. The bottom row of the block contains the outputs of the calculation and nothing else. Only the bottom row of a calculation block should be referenced by other cells.

The technique of using of "daisy chain" cell references is avoided by always referencing the original source directly. This facilitates deletion of unneeded calculation blocks without risking the integrity of later calculations, supporting specification changes and maintenance.

Using the FAST approach requires writing code in a large number of rows in the spreadsheet. Although construction techniques are not part of the standard, to build efficiently this large number of rows it is, in our view, highly advantageous that the programmer have skill (or "muscle memory") with certain keyboard shortcuts and certain facilitating practices to make routine operations very quick and accurate. These keyboard shortcuts have several benefits: they avoid fine-motor control mouse actions; eliminate delays from shifting hands between keyboard and mouse; allow writing a cell formula once then copying it; allow for entire rows to be copy and pasted; and enable easy navigation across sections of a worksheet. Facilitating practices include careful, consistent use of absolute references and hiding unused columns to the right of the model. The benefits seem analogous to a writer learning to touch-type rather than "hunt-and-peck". In our judgment, without keyboard shortcut skills many programmers will be tempted to skip some of the FAST design standards in the interest of short-term time savings, and therefore training in these skills is a practical necessity.

## 5. THE OPERIS SPREADSHEET ENGINEERING METHODOLOGY

The Operis methodology is used and taught by Operis Group plc. Operis states that they train over 1,000 analysts per annum. The Operis practices are documented in Swan (2008), a subtle and lucid book with accompanying Excel files that provides extensive recommendations on design and programming practices.

The Operis approach is focused on financial modelling[1]. It recommends complete specification of the outputs prior to the start of coding (see Swan 2008, Chapter 1). Each output is provided on its own worksheet and is crafted to be effective when viewed or printed. The example model we saw had nine outputs; an executive summary worksheet plus 8 worksheets with financial statements. Benefits of this approach include focus on end goals prior to coding, facilitation of early feedback, and ability to print the current version on short notice.

The analyst starts with a blank workbook and constructs any number of output worksheets. The output worksheets are programmed initially as a "skeleton" containing column headers, extensive row labels, no numbers, and but a few simple formulas for basic finance concepts. The analyst then constructs a single "Inputs" worksheet and a single "Workings" worksheet. Every worksheet has a common column layout featuring year headers on the rightmost columns.

We perceive that the Operis approach requires in the pre-coding phase great discipline, substantial analyst sophistication with financial concepts, dialog with the client, and perhaps some level of client sophistication or trust in the analyst. Although this is different than our usual experience with spreadsheet developers, we note that this approach is entirely consistent with the recommendations from traditional software engineering (for example, McConnell 2004).

The spreadsheet is constructed in a disciplined, assembly line fashion by adding and completing one "model feature" (our term) at a time. A model feature is one calculation (or a handful of intertwined calculations) with associated inputs and outputs. The model feature is added as a module on the Inputs worksheet, then the Workings worksheet, then the output worksheets. On the Inputs and Workings worksheets, the new module is entered just below the previously-entered module, separated by a blank row. On the appropriate output worksheet(s), the module is entered into the appropriate pre-labeled location.

---

[1] Operis (2010a) raises the distinction between "mode_ll_ing" and "mode_l_ing". There is no agreement on the use of these two terms. We use "mode_l_ing" to refer to a conceptual model, and "mode_ll_ing" to refer to a spreadsheet representation of a conceptual model.

Each module is programmed step-by-step. On the Inputs worksheet a new row label is added and given a Name documented at the end of the row, and data is entered. On the Workings worksheet, a new row label is added. Cell formulas are programmed in the row. Cell formulas are kept short. The row is given a Name, documented at the end of the row. Each row of new data from the Inputs worksheet is linked (i.e., referenced) in the new module on the Workings worksheet. The Inputs worksheet is referenced only from the Workings worksheet. On each output worksheet that needs the new information, links are written to the appropriate new row(s) in the Workings worksheet. An output worksheet references only the Workings worksheet and perhaps itself.

The Operis programming approach recommends (but does not require) Excel Names. Operis consistently defines Names by row, where a range Name refers to all values across the time column. The approach relies upon a subtle and lesser-known Excel "column matching" feature. (Note that this feature works with Names and with coordinate notation.) Suppose you define a Name called "Cost" that refers to cells $G$8:$K$8, and write in cell H12 the formula =Cost. Excel uses "column matching" to return in cell H12 the value from cell H8 (see Swan 2008, p. 90). This "column matching" feature works even when you write the formula =Cost in cell H12 *on a different worksheet*. If you write =Cost in a cell outside the defined columns G:K, Excel returns a #VALUE! error. This approach requires ruthless column standardization across worksheets.

This approach would seem to eliminate the possibility of referencing an incorrect column on a different worksheet. It enables an absolute reference cell formula to be written once and copied across, so formulas are *identical* as columns are traversed rather than incremented, greatly increasing readability. (Some traditional non-identical formulas still occur such as period-to-period inflators and "corkscrew" formulas.) There is a proliferation of names, but with disciplined development this need not be burdensome. Automation in the form of keyboard shortcuts are used heavily.

The Operis Analysis Kit (OAK) add-in (Operis 2010b) provides powerful automation and productivity features, including: Insert rows and columns (even when they intersect merged cells, arrays and data tables); Get suggestions for optimizing formulas; Rich options for managing Excel names including changing names and removing names; Overview of a new spreadsheet that includes a list of all the worksheets in a workbook, a sortable list of every formula that is distinct, a list of the constants, a diagram showing a bird's eye view of the patterns in each worksheet, a sortable list of all the names in a workbook; Formula "deciphering" that includes reconstruct a formula to receive automated re-performance of specified calculations and understand which cells they act on, and prune a formula to make a calculation easier to understand by showing only the elements that are actually active at the current moment, identify the discrepancies that are preventing your attempt at a parallel reconstruction of a spreadsheet calculation from reproducing the original accurately; Spreadsheet comparison that includes distinguishing changes that are real from ones that are merely the consequence of items changing position, systematic version control, and other features.

The Operis approach has provisions, some subtle, for an extensive Audit worksheet for quality control, and a method for handling scenarios by toggling between different input worksheets.

# 6. THE SSRB SPREADSHEET ENGINEERING METHODOLOGY

Best Practice Spreadsheet Modelling Standards (SSRB 2005a) is promulgated by the Spreadsheet Standards Review Board (SSRB). The SSRB was established by BPM Analytical Empowerment Pty Ltd. SSRB is described as an independent body of members with expertise and reputation in developing and using spreadsheets. However, we cannot find any listing of members or contributors. The SSRB address and phone number are the same as BPM Analytical's headquarters. At time of writing, we perceive SSRB as a subsidiary of BPM Analytical rather than a multi-party "review board"; any claim to being an independent organization should, in our view, be considered aspirational.

The Standards were first published in July 2003 and the Standard that is currently available is version 4.1 published in October 2005. Copyright is owned by BPM Analytical. (Note: On May 24, 2010 BPM Analytical released version 6.0 of the bpmToolbox and SSRB released version 6.0 of the Best Practice Modelling Standard. This release occurred during the revision round of this paper and we are unable to review the new versions here. There seem to be substantial enhancements over the previous versions.)

SSRB views spreadsheets as the primary vehicle for modelling in business and argue that lack of standardization in spreadsheet development process leads to frustration, confusion and time lost. They recommend that there be universally applicable, freely available and definitive principles governing the spreadsheet model development process. The most significant benefits of the consistent use of such principles are quoted as Improved transparency; Reduction in development time and cost; Reduction of error rates.

SSRB distinguishes between Modeling "Standards" and "Conventions". A standard is in some sense required whereas a convention is recommended. For example, different type of cells are required to be colored differently as a "Standard", the color choice for , say input cells, is recommended as a "Convention".

The standards are carefully organized into 16 "Spreadsheet Modelling Areas" each of which contains two to nine subsections. Key recommendations from the standards document include modularization by workbook and worksheet. Similar worksheets should be grouped into a "section" and delimited by a cover worksheet. The standard is silent about modularization within worksheets, but it does indicate that worksheets should be only a single screen and that larger worksheets should be split into multiple worksheets, suggesting that the smallest unit of modularization in the worksheet. SSRB pays close attention to column structure in financial analysis.

SSRB recommends documentation at all possible levels in varying forms from inclusion of cover worksheets to the use of descriptive labels and Excel Names. They argue for entering data once and referencing it, and for protection of all non-input cells.

SSRB accepts complex cell formulas but suggests that each function be listed on a different line in the formula bar. It distinguishes among "summary view (compacted), "print view (semi-compacted, if required)" and "expanded view (un-compacted)". We are not sure what compaction means but it might refer to concealing rows and columns via hiding or grouping.

The recommended practices become very detailed, with rules on what font/fill options to use for different type of cells (for example, the use of blue font for constants), rules on what specific labels to use while naming worksheets, and rules on how to deal with complex formulae. An interesting feature of the Standards document is that it pays particular attention to how to prepare sheets for time series analysis – mainly designated for financial models that operate over multiple periods.

The SSRB standard has a separate document (SSRB 2005b) that discusses how to convert an arbitrary spreadsheet to the SSRB standard. It shows reliance upon an Excel add-in to manage the creation of worksheets. This add-in may have many other features but we could not find details on the website.

BPM Analytical Empowerment Pty Ltd. publishes a powerful spreadsheet development productivity add-in called bpmToolbox (BpmToolbox 2010) that automates the execution of many common construction and editing tasks, including: Best practice formats & styles; Base sheets system; 'Quick' efficiency tools; Table of contents; Workbook restructuring; Multiple workbook tools; Keys; Time series analysis ;Automated error checks; Content creation tools; Finalization & review tools; Theme customization; and Theme creation and sharing. BpmToolbox has strong integration with SSRB standards and conventions. Although an adequate analysis is outside the scope of this paper, we hypothesize that use of bpmToolbox may provide a very substantial increase in spreadsheet programmer productivity.

## 7. COMPARE/CONTRAST OF THE STANDARDS

To compare the three methodologies we use the nine-activity spreadsheet engineering paradigm proposed by Grossman and Özlük (2004).

### 7.1. MODELING

*Modeling* is the process of determining what the spreadsheet is to do. This can be thought of as a specifications statement, or as defining the outputs prior to the start of programming.

Although the mention of sensitivity analysis and time series analysis by the SSRB standards can be considered part of modeling (particularly from an operations research perspective), there is no direct reference to model specification process. FAST does not explicitly discuss any topics immediately pertaining to modeling. Operis requires a complete, detailed output specification.

### 7.2. DEVELOPMENT PARAMETERS

*Development parameters* are the planning assumptions of a spreadsheet, including goal, budget, users, usage, and lifecycle intentions. All three methodologies have the goal of creating spreadsheets in the domain of financial planning models. They all support building spreadsheet applications that will be deployed to one or more non-programmer users that can accept different input data (or "scenarios", which are sets of input data), with a lifecycle that includes future modification.

### 7.3. DESIGN

*Design* comprises the way cells are arranged (structural design) and the appearance of cells and cell borders (visual design).

Two principles of structural design are widespread. The first is to organize related concepts using the rows and columns of the spreadsheet. FAST standards specifically ask the spreadsheet developers to maintain a consistent column/row structure across all sheets to improve readability.

They also recommend that each column within a sheet have a single and consistent purpose. SSRB standards do not provide a clear set of recommendations on the issue. The second structural design principle is "modularity", which says that logically related elements be grouped into modules. Modularity gets a lot of traction both in the FAST standards and the SSRB standards. They strongly argue that worksheets should be separated by functionality such as input worksheets, presentation worksheets etc.; there should also be clear blocks with separate functionality within each worksheet whenever necessary.

Visual design is essentially the formats applied to the spreadsheet. The most important visual design elements are the use of fonts (including font choice, color, and bold/italic), justification within a cell, cell colors, and cell borders. FAST and SSRB standards spend a considerable amount of time laying out recommendation about the visual elements of spreadsheet design. SSRB standards require that the function of each sheet, block within a sheet and each cell be visually identifiable through appropriate use of visual elements such as font sizes, colors, fill patterns etc. SSRB has one section entirely devoted to describing format and style requirements. FAST is also very clear at asking for clear indications for cells types; in some cases, it goes as detailed as to prescribe specific colors for imported values and values to be exported. In addition, FAST advices that sheets are arranged so that information flows left to right.

### 7.4. PROGRAMMING

*Programming* is the creation of cell formulas and other logic in a spreadsheet. There is agreement on what we think of as "good hygiene". Data is entered once and referenced. Avoid constants in formulas. No circular references. Formulas should be the same across a row (sometimes called "left to right consistency"). Use a modular design.

All three methodologies recognize that cell formula length needs to be managed. FAST and Operis dictate short formulas. (FAST employs an a rule of thumb, namely "a formula longer than a thumb is likely to be too long".) SSRB is less intolerant of long cell formulas but suggests that each function appear on a separate line in the Formula bar.

FAST and Operis have rigorous procedures for writing programming statements. FAST and Operis construct a program in a similar fashion. One small module is written at a time. The inputs are carefully referenced in a standard way. FAST always brings a copy of the inputs to the module so they are adjacent and visible. Operis bring the inputs to the Workings sheet the first time the input is used. Both methodologies write a cell formula in a far left column, then copy it across to the last column. The act of programming is standardized, mechanical and quick. We infer from the SSRB materials on their website that they may have a similar process, but are not certain.

There is disagreement on the use of Excel Names. The main difficulty with Names is that as proliferate they become long. Long Names can make cell formulas hard to read and are hard to manage. Therefore, FAST uses Names only for a handful of global variables. Operis makes extensive, systematic use of Names. Their approach of using row Names and the Name/column intersection is a reasonable solution to the difficulties described above. SSRB recommends heavy reliance on Names and provides a means to create standard, meaningful names.

### 7.5. QUALITY CONTROL

*Quality Control* is all actions taken to determine whether the outputs of a spreadsheet are satisfactory. It is well-known in traditional software engineering that the best way to insure quality is to "do it right the first time" by not writing errors, and that the best time to detect an error is immediately after you have created it.

FAST, Operis and SSRB have rigorous coding practices that, in our view, greatly reduce the likelihood of writing erroneous cell formulas compared to *ad hoc* development practices. FAST, Operis and SSRB all have provisions for checks on errors and anomalies of various types and use them extensively during the programming process. The details are outside the scope of this paper.

For some types of software, a powerful and widely-used quality control technique is testing. Testing performs extensive variations on the inputs to the software to confirm that the proper outputs are generated. Testing requires that the QC group has the ability to determine the correct outputs for any set of inputs. In the event that the correct outputs are not knowable, testing is of little value. For this reason software testing is not applicable to a large class of scientific and business models, including large financial planning models, because the only calculation of the outputs is computed by the software being tested. Computer science has not generated a satisfactory solution to this challenge; techniques include having two independent teams develop software and then compare the results, and performing a manual review with a calculator. Both these techniques consume excessive time and resources and are not widely used.

### 7.6.  DEBUGGING

*Debugging* is modifying a spreadsheet program to fix an output that has an unsatisfactory [incorrect] value. None of the methodologies provides recommendations for finding root causes and fixing them correctly.

SSRB has a methodology for converting other spreadsheets to its standard, which is could be considered a way of surfacing and fixing bugs. Operis has told us that they have a business of taking existing spreadsheets and reprogramming them to their standard, which surfaces errors in the original spreadsheet and allows easy remediation. F1F9 has told us they the same.

This implies that the methodologies support, at least implicitly, a debugging technique of reprogramming a suspect model according to one's preferred standard. Such a technique requires the analyst to have strong financial knowledge; more generally, the analysts must have strong domain knowledge.

### 7.7.  DOCUMENTATION

*Documentation* is any written record regarding the spreadsheet. All the methodologies rely on standard, consistent, rigorous use of row and column labels. They have very similar use of column labels. FAST and Operis clearly indicate Names adjacent to named ranges whereas SSRB suggests use of specific prefixes for range names to distinguish different features of what the name refers to. All three have a combination of programming clarity, right to left consistency, and labels that makes the code largely self-documenting. SSRB has a higher level of "explanatory" documentation in the form of cover worksheet describing basic information about the spreadsheet model and the developer team. Similarly, FAST advises that the model builder develop a control sheet to provide a table of contents and a list of model qualifications and weaknesses.

### 7.8. USAGE

*Usage* of a spreadsheet is any process where a user provides inputs to a spreadsheet, and observes the outputs. All three methodologies support repeated usage by a user who is not the programmer.

### 7.9. MODIFICATION

*Modification* refers to changes made to the spreadsheet after it has been used. This includes terms such as "maintenance", "enhancement" and "extension". This is essential for spreadsheets that will be used over time. The methodologies do not explicitly address modification. However, we believe they were written with modification in mind.

All three methodologies create highly readable well-documented code in a standardized, repeatable way. We believe that other analysts trained in that methodology should be able efficiently to modify the code, at least as long as the modifications are within the domain of financial planning models. In our interaction with F1F9 staff, we were able to observe how the FAST methodology clearly supports maintenance.

## 8. CONCLUSIONS

### 8.1. Caveats

The FAST, Operis and SSRB methodologies each comprise a rich integrated set of practices and procedures. In the short summaries provided in this paper we can only hit the most essential elements and cannot fully communicate their richness and power.

We learned in doing this research that the standalone written materials for the FAST and Operis methodologies were not sufficient for us to adequately grasp their most important aspects. We doubt we would have fully appreciated important subtleties without access to training materials and personal communication. We are able to call upon prior relationships or visits to fill in the gaps before the submission deadline for this paper. Regrettably we were unable to do so with SSRB. We acknowledge that our understanding of the SSRB methodology is less we would like.

### 8.2. Differences in Terminology

There is no standard terminology for building spreadsheet models. Different methodologies use terms specific to their approach. Different communities outside of FAST, Operis and SSRB use their own terms. We note that SSRB and FAST provide a glossary of terms.

This confusion in terminology makes it harder to communicate about spreadsheet development. Here are a few examples. Terms such as "reference", "link", "echo" and "point" are used to describe a cell formula that references only a single cell, for example "=B6", or perhaps a single range of cells. "Link" can also refer to hyperlinks that reference a cell in a spreadsheet or a hyperlink to an external resource. Some developers use terms such as "anchor", "row anchor" for absolute vs. relative cell referencing; for example an 'absolute cell reference' is referred to as being 'fully anchored' in FAST terminology.

In the language of operations research modeling, quality control for a computer model has two components. First, the model must be "validated" to insure that it is meaningful. For spreadsheet financial models this corresponds to confirming the commercial logic is correct. Second, the computer implementation of the model needs to be "verified" to insure that it is coded correctly. For spreadsheet financial models this corresponds to confirming that there are no programming errors. Terminology associated with quality control includes "verify", "review", "audit", "inspect", and "check". These words can have different meanings in different contexts.

### 8.3. Dissemination of Methodologies

We believe that, at least at their current level of development, these methodologies will most effectively be disseminated via training courses rather than documents such as books and standards. This is analogous to the field of six sigma quality control, where countless books have been written yet attendance at courses is recognized as essential because of the benefit of obtaining hands-on experience under the tutelage of an expert.

### 8.4. These Spreadsheet Engineering Methodologies Have Narrow Domain

We emphasize that the FAST, Operis, and SSRB spreadsheet engineering methodologies have tightly defined domains. They do not attempt to address "writing spreadsheets" in general. They do not attempt to address "writing finance spreadsheets" because these approaches would not be suitable for finance models such as investment portfolio optimization. These methodologies apply to the domain of large financial planning models. In addition, the methodologies seem to be used by a programming team with solid domain knowledge so the nature and computation of the outputs is broadly understood.

### 8.5. The Professionalization of Spreadsheet Programming

In a classic computer programming book Weinberg (1998, p. 125) draws a distinction between the amateur and professional programmer. The amateur wants to learn about his area of domain expertise, whereas the professional wants to learn about programming.

> "The amateur [programmer], being committed to the results of the particular program for his own purposes, is looking for a way to get the job done. If he runs into difficulty, all he wants is to surmount it—the manner of doing so is of little consequence."

> "Not so, however, for the professional [programmer]. He may well be aware of numerous ways of circumnavigating the problem at hand…But his work does not stop there; it begins there. It begins because he must understand why he did not understand, in order that he may prepare himself for the programs he may someday write which will require that understanding."

The three spreadsheet engineering methodologies discussed in this paper clearly represent the perspective of the *professional programmer*, who has invested much effort over a long period of time to "prepare himself for the programs he may someday write". The three methodologies provide a solution to the "spreadsheet engineering deployment research issues" identified by Grossman (2002), including professionalization and strict, centralized development policies.

The FAST, Operis, and SSRB methodologies are clear examples of the professionalization of spreadsheet programming. Like any sound methodology, they apply to a limited domain - that of large financial planning models. These methodologies may serve as a model for other domains.

### 8.6. Spreadsheet Verification and Debugging

The important task of verification (sometimes called "audit" or "inspection") of a spreadsheet to insure the commercial logic is correct and the spreadsheet implementation of the commercial logic is accurate remains an important challenge. Once a verification process identifies an error there is then the issue of how to fix it without introducing new errors. Despite many products for spreadsheet checking and auditing (see Spreadsheet Analytics 2010), there is no accepted solution for these problems.

Several people we contacted during this research indicate they use a powerful technique for verification and debugging. The technique is to simply rewrite the spreadsheet model using one's preferred spreadsheet engineering methodology. Use of this technique requires knowledge of the applicable methodology, some additional spreadsheet programming skills (undoubtedly standardized or standardizable) and knowledge of the model's domain.

### 8.7. How Effective Are These Methodologies?

Evaluating the effectiveness of a spreadsheet engineering methodology is challenging. None of the methodologies have received rigorous third party evaluation regarding accuracy and programmer productivity. So for the reader demanding solid proof, there is none. However, there are three reasons to find credible the assertion that these methodologies provide benefits: First is that software engineering provides benefits; second is vendors' self-reported experience with reprogramming extant spreadsheets; and third is the authors' direct experience.

All three methodologies employ accepted software engineering principles that underlie software engineering methodologies for third-generation programming languages (for example, see McConnell 2004). There is an extensive literature showing that use of such methodologies leads to improvements in productivity, accuracy, maintainability, and cost. We would expect that software engineering methodologies relying on the same principles and tailored to spreadsheets would have similar benefits.

FAST and Operis representatives were able to explain their experience in reprogramming extant spreadsheets using their methodology. (SSRB also has a methodology for making such a conversion.) They tell us that this experience, which surfaces differences between the original spreadsheet and the reprogrammed spreadsheet, allows comparison of both spreadsheets; the result almost invariably is that the difference arises from an error in the original spreadsheet.

The authors have extensive experience building spreadsheets and observing how business students build spreadsheets before and after training in spreadsheet engineering principles and practices. Based on this experience, we found that the OAK and FAST methodologies provided, for us, dramatic improvement in efficiency and the elimination of many repetitious and other error-prone activities, including activities that we had long perceived as problematic. We reached similar conclusions based on observation of the SSRB methodology with the bpmToolbox add-in.

Therefore, in our judgment, we find credible the assertion that careful use of the FAST, Operis, or SSRB spreadsheet engineering methodology will lead to enhanced productivity, accuracy, and maintainability of large financial spreadsheet models.

## 8.8. Future Research

It would be interesting to evaluate one or more spreadsheet engineering methodologies in light of traditional software engineering practices. In particular, one could evaluate a spreadsheet engineering methodology and the organization that uses it with the well-established Capability Maturity Model (CMMI 2010). The efficiency and effectiveness of the verification approach of rewriting the spreadsheet model discussed in section 8.6 above merits study.

## ACKNOWLEDGMENTS

We thank Morten Siersted, Dhruva Poonia, other F1F9 staffers, and John Richter for their assistance in understanding the FAST methodology. We thank David Colver and Jonathan Swan for their assistance in understanding the Operis methodology. We thank Michael Hutchens for his assistance in understanding the SSRB methodology.

All errors of omission or commission are the sole responsibility of the authors.

## REFERENCES

bpmToolbox (2010), "Introducing bpmToolbox", http://www.bestpracticemodelling.com/software/bpmtoolbox, accessed June 6, 2010.

CMMI (2010), http://www.sei.cmu.edu/cmmi/, accessed March 31, 2010.

FAST (2010), *FAST Modeling Standard: Practical, structured design rules for financial modeling, Version FAST01a*, http://www.fast-standard.org/, accessed March 31, 2010.

Grossman, T. A. (2002), "Spreadsheet Engineering: A Research Framework", *European Spreadsheet Risks Interest Group 3rd Annual Symposium*, Cardiff, Wales, July.

Grossman, T. A. (2006), "Integrating Spreadsheet Engineering in a Management Science Course--A Hierarchical Approach", *INFORMS Transactions on Education* 7(1), pp. 18-36, September, available at http://www.informs.org/Journal/ITE/Archive/Volume-7/Integrating-Spreadsheet-Engineering-in-a-Management-Science-Course-A-Hierarchical-Approach.

Grossman, T. A., V. Mehrotra, and Ö. Özlük (2007), "Lessons from Mission-Critical Spreadsheets", *Communications of the Association for Information Systems* 20(60), pp. 1009-1042, December.

Grossman, T. A. and Özlük, Ö. (2004), "A Paradigm for Spreadsheet Engineering Methodologies", *Proceedings of the European Spreadsheet Risks Interest Group 5th Annual Symposium*, Klagenfurt, Austria, July.

McConnell, S. (2004), *Code Complete, 2$^{nd}$ Edition*, Microsoft Press.

Operis (2010a), "Operis Analysis Kit help feature: Financial Modelling or Modeling", http://www.operisanalysiskit.com/help/2007/index.html?internationalspelling.htm, accessed May 10, 2010.

Operis (2010b), "About OAK", http://www.operisanalysiskit.com/AboutOak.htm, accessed June 6, 2010.

Powell, S. and Baker, K. (2009). *Management Science: The Art of Modeling with Spreadsheets 2$^{nd}$ edition*, John Wiley & Sons.

Raffensperger, J. F. (2001), "New Guidelines for Spreadsheets", *Proceedings of the European Spreadsheet Risks Interest Group*, Amsterdam, pp. 61-76, July, available at http://arxiv4.library.cornell.edu/ftp/arxiv/papers/0807/0807.3186.pdf, accessed March 29, 2010.

Read, N. and J. Batson (1999), "Spreadsheet modeling best practice", http://www.eusprig.org/smbp.pdf, accessed May 31, 2010.

Spreadsheet Analytics (2010), "Checking & Auditing", in *Spreadsheet Analytics: Resources for Spreadsheet Analysts*, http://www.usfca.edu/bps/spreadsheet-analytics/development-management/checking-auditing, accessed May 10, 2010.

SSRB (2005a), *Best Practice Spreadsheet Modeling Standards, version 4.1* available at http://www.ssrb.org/, accessed March 29, 2010.

SSRB (2005b) *Converting Excel workbooks to Best Practice Spreadsheet Models, version 3.1*, available at http://www.ssrb.org/best_practice_spreadsheet_modelling_standards_support.html, accessed March 29, 2010.

Swan, J. (2008), *Practical Financial Modelling: A Guide to Current Practice, 2$^{nd}$ Edition*, CIMA Pubishing.

Tennent, J., G. Friend, (2001), *Guide to Business Modelling*, Profile Books, London.

Weinberg, G. (1998), *The Psychology of Computer Programming, Silver Anniversary Edition*, Dorset House.